\documentclass[reprint,onecolumn]{revtex4}%
\usepackage{amsfonts}
\usepackage{amsmath}
\usepackage{amssymb}
\usepackage{charter}
\usepackage{graphicx}%
\providecommand{\U}[1]{\protect \rule{.1in}{.1in}}
\begin{document}
\title{Intermediate Coherent-entangled State Representation: Generation and its applications}
\author{Qin Guo$^{1,2,3,\dag}\thanks{Corresponding author: guoqin91@163.com}$, Li-ying
Jiang$^1$, Cun-jin Liu$^{1,2,3}$, Ming Cai$^4$ and Li-yun
Hu$^{1,2,3,\dag}
\thanks{Corresponding author: hlyun2008@126.com}$}
\affiliation{$^{1}$College of Physics and Communication Electronics, Jiangxi Normal
University, Nanchang 330022, China}
\affiliation{$^{2}$Center for Quantum Science and Technology, Jiangxi Normal University,
Nanchang 330022, China}
\affiliation{$^{3}$Key Laboratory of Optoelectronic and Telecommunication of Jiangxi,
Nanchang 330022, China}
\affiliation{$^{4}$Library, Jiangxi Normal University, Nanchang 330022, China }

\begin{abstract}
By combining the beam splitter and the Fresnel transform, a protocol is
proposed to generate a new entangled state representation, called the
intermediate coherent-entangled state (ICES) representation. The properties,
such as eigenvalue equation, completeness relation and orthogonal relation,
are investigated. The conjugate state representation of the ICES and the
Schmidt decomposing of the ICES are also discussed. As applications, a new
squeezing operator and some operator identities by using the ICES are obtained.

\end{abstract}
\maketitle

\section{ Introduction}

The quantum mechanical representation theory proposed by Dirac \cite{1} has
made a profound effect on the development of modern quantum physics. The
symbolic method combined with the technique of integration within an ordered
product (IWOP) of operators has been found to be a usefull method to find the
new quantum mechanical representation \cite{2,3,4}. In recent years, entangled
state representations \cite{5} have played a more important role in quantum
optics \cite{6,7}, quantum information and quantum computation
\cite{8,9,10,11}.

Actually, quantum mechanical representation can construct a bridge between
classical linear integral transform and quantum unitary operator \cite{11a}.
That is to say, different linear integral transforms can be described by
different quantum operators, such as the generalized Fresnel transform
\cite{12} in 1-dimensional (1D) can be described by the 1-mode Fresnel
operator $F_{1}\left(  r,s\right)  $ \cite{13} and the 2-mode Fresnel
transform can be described by 2-mode Fresnel operator $F_{2}\left(
r,s\right)  $ \cite{14} in quantum optics. These quantum operators acting on
quantum states can generate new quantum mechanical representations. For
instance, Fresnel operator $F_{1}\left(  r,s\right)  $ acting on the
coordinate eigenstate $\left \vert q\right \rangle $ can generate the
intermediate coordinate-momentum representation $\left \vert q\right \rangle
_{s,r}$ \cite{15}, and Fresnel operator $F_{2}\left(  r,s\right)  $ applying
to the entangled state $\left \vert \eta \right \rangle $ \cite{16} can generate
the intermediate entangled state representation $\left \vert \eta \right \rangle
_{s,r}$ \cite{2,15}.

On the other hand, beam splitter (BS) has played an important role in
generating entangled states, such as a three-mode continuous-variables (CV)
entangled state was proposed by using an asymmetric BS \cite{17,18,19,20} and
a parametric down-conversion (PDC) instrument, and then a four-mode CV
entangled state was generated by BS and PDCs \cite{21}. In addition,
multi-mode CV entangled state was generated by a BS and a polarizer or by BS,
PDC and a polarizer \cite{18,22,24}. It is no doubt that all these studies
above exhibit the role of BS in generating entangled states and indicate that
these physical instruments, such as PDC, BS and polarizer, can be used to
implement entangled states.

Then an interesting question thus naturally arises: is there any kind of
entangled state which can be generated by combining BS and Fresnel transform
optical process? As far as our knowledge goes, there has no report in the
literature up to now. In this paper, we will introduce a new entangled state
representation which is called the intermediate coherent-entangled state
(ICES) representation. It is shown that it can be generated by the BS and the
Fresnel transform.

The paper\ is arranged as follows. In Section 2, we introduce the Fresnel
operator and the intermediate coordinate-momentum state (ICMS) representations
which can be obtained by the Fresnel operator \cite{11a}. In section 3, we
present two methods to produce ICES, which is a new entangled state
representation and is not been presented before. Deriving the ICES
representation naturally from the completeness relations of the coherent state
and the generalized coordinate state by IWOP technique is one method.
Producing the ICES by combining a symmetric BS and the Fresnel transformation
is another method. In section 4, we discuss the characters of the ICES,
including the eigenvalue equation, orthogonal relation, Schmidt decomposition
and conjugate state. As the applications of the new representation, in section
5, we derive a new squeezing operator and derive some operator identities. We
end our work in section 6. These discussions exhibit not only the usefulness
of the ICES representation, but also provide a considerable insight into the
character of them.

\section{Fresnel operator and the ICMS}

In Ref.\cite{15,26,27,28,29}, the Fresnel operator $F_{1}\left(  r,s\right)  $
is proposed by using the coherent state representation $\left \vert
z\right \rangle =\exp \left[  -\frac{1}{2}\left \vert z\right \vert ^{2}%
+za^{\dagger}\right]  \left \vert 0\right \rangle $ and the IWOP technique
\cite{2}, which is given by
\begin{equation}
F_{1}\left(  r,s\right)  =\frac{1}{\sqrt{s^{\ast}}}\exp \left(  \frac
{r}{-2s^{\ast}}a^{\dagger2}\right)  \colon \exp \left[  \left(  \frac{1}%
{s^{\ast}}-1\right)  a^{\dagger}a\right]  \colon \exp \left(  \frac{r^{\ast}%
}{2s^{\ast}}a^{2}\right)  , \label{1}%
\end{equation}
where the symbol $\colon \colon$ denotes the normally ordering, and $s$ and $r$
are complex and satisfy the unimodularity condition $ss^{\ast}-rr^{\ast}=1$.
The Fresnel operator $F_{1}\left(  r,s\right)  $ corresponds to optical
Fresnel transformation characteristic of ray transfer matrix elements $\left(
A,B,C,D\right)  $, $AD-BC=1$, connecting the input light field $f(x)$ and
output light field $g(x)$ by the Fresnel integration \cite{14},
\begin{equation}
g\left(  x^{\prime}\right)  =\frac{1}{\sqrt{2\pi iB}}\int_{-\infty}^{\infty
}\exp \left[  \frac{i}{2B}\left(  Ax^{2}-2x^{\prime}x+Dx^{\prime2}\right)
\right]  f\left(  x\right)  dx. \label{2}%
\end{equation}
Parameters $r$ and $s$ in Eq.(\ref{1}) are related to $\left(
\begin{array}
[c]{cc}%
A & B\\
C & D
\end{array}
\right)  $ transfer matrix by%
\begin{align}
s  &  =\frac{1}{2}\left[  \left(  A+D\right)  -i\left(  B-C\right)  \right]
,\nonumber \\
r  &  =-\frac{1}{2}\left[  \left(  A-D\right)  +i\left(  B+C\right)  \right]
. \label{3}%
\end{align}
Let the Fresnel operator $F_{1}\left(  r,s\right)  $ act on the coordinate
eigenstate $\left \vert q\right \rangle $, and use the completeness relation of
the coherent state $\int \frac{d^{2}z}{\pi}\left \vert z\right \rangle
\left \langle z\right \vert =1$, we obtain the ICMS $\left \vert q\right \rangle
_{s,r}$ \cite{15},
\begin{align}
&  F_{1}\left(  r,s\right)  \left \vert q\right \rangle \nonumber \\
&  =\frac{\pi^{-1/4}}{\sqrt{s^{\ast}+r^{\ast}}}\exp \left[  \frac{r^{\ast
}-s^{\ast}}{2\left(  s^{\ast}+r^{\ast}\right)  }q^{2}+\frac{\sqrt{2}q}%
{s^{\ast}+r^{\ast}}a^{\dagger}-\frac{\left(  s+r\right)  }{2\left(  s^{\ast
}+r^{\ast}\right)  }a^{\dagger2}\right]  \left \vert 0\right \rangle \nonumber \\
&  \equiv \left \vert q\right \rangle _{s,r}, \label{4}%
\end{align}
where we used the following relations,%
\begin{equation}
\left.  \left \langle z\right \vert q\right \rangle =\pi^{-1/4}\exp \left \{
-\frac{q^{2}}{2}-\frac{\left \vert z\right \vert ^{2}}{2}+\sqrt{2}qz^{\ast
}-\frac{z^{\ast2}}{2}\right \}  , \label{5}%
\end{equation}
and the integration formula%
\begin{align}
&  \int \frac{d^{2}z}{\pi}\exp \left(  \zeta \left \vert z\right \vert ^{2}+\xi
z+\eta z^{\ast}+fz^{2}+gz^{\ast2}\right) \nonumber \\
&  =\frac{1}{\sqrt{\zeta^{2}-4fg}}\exp \left[  \frac{-\zeta \xi \eta+\xi
^{2}g+\eta^{2}f}{\zeta^{2}-4fg}\right]  , \label{6}%
\end{align}
whose convergent condition is Re$\left(  \zeta \pm f\pm g\right)
<0,\ $Re$\left[  \left(  \zeta^{2}-4fg\right)  /\left(  \zeta \pm f\pm
g\right)  \right]  <0$. According to Eq.(\ref{3}), we also have
\begin{equation}
\left \vert q\right \rangle _{s,r}=\frac{\pi^{-1/4}}{\sqrt{D+iB}}\exp \left[
-\frac{A-iC}{D+iB}\frac{q^{2}}{2}+\frac{\sqrt{2}q}{D+iB}a^{\dagger}%
-\frac{D-iB}{D+iB}\frac{a^{\dagger2}}{2}\right]  \left \vert 0\right \rangle .
\label{7}%
\end{equation}
Correspondingly, the eigenvalue equation of $\left \vert q\right \rangle _{s,r}$
is%
\begin{equation}
\left(  DQ-BP\right)  \left \vert q\right \rangle _{s,r}=q\left \vert
q\right \rangle _{s,r}, \label{8}%
\end{equation}
where $Q=(a+a^{\dagger})/\sqrt{2},P=(a-a^{^{\dagger}})/(i\sqrt{2})$. When
$A=D=1,B=C=0,$the above equation evolves into $Q\left \vert q\right \rangle
_{s,r}=q\left \vert q\right \rangle _{s,r}$, and $\left \vert q\right \rangle
_{s,r}$ is just the coordinate operator eigenstate $\left \vert q\right \rangle
$. When $A=D=0$, $-B=C=1$, the above equation evolves into $P\left \vert
q\right \rangle _{s,r}=q\left \vert q\right \rangle _{s,r},$and $\left \vert
q\right \rangle _{s,r}$ is just the momentum operator eigenstate $\left \vert
p=q\right \rangle $. So $\left \vert q\right \rangle _{s,r}$ is named
intermediate coordinate-momentum state.

In a similar way, operating the Fresnel operator on the momentum eigenstate
$\left \vert p\right \rangle $, together with the completeness relation of the
coherent state, we obtain the intermediate momentum-coordinate state (IMCS)
$\left \vert p\right \rangle _{s,r}$,%

\begin{align}
&  F_{1}\left(  r,s\right)  \left \vert p\right \rangle \nonumber \\
&  =\frac{\pi^{-1/4}}{\sqrt{s^{\ast}-r^{\ast}}}\exp \left[  \frac{\sqrt{2}%
ip}{s^{\ast}-r^{\ast}}a^{\dagger}+\frac{s-r}{2\left(  s^{\ast}-r^{\ast
}\right)  }a^{\dagger2}-\frac{s^{\ast}+r^{\ast}}{2\left(  s^{\ast}-r^{\ast
}\right)  }p^{2}\right]  \left \vert 0\right \rangle \nonumber \\
&  \equiv \left \vert p\right \rangle _{s,r}, \label{9}%
\end{align}
or%
\begin{equation}
\left \vert p\right \rangle _{s,r}=\frac{\pi^{-1/4}}{\sqrt{A-iC}}\exp \left[
-\frac{D+iB}{A-iC}\frac{p^{2}}{2}+\frac{\sqrt{2}ip}{A-iC}a^{\dagger}%
+\frac{A+iC}{A-iC}\frac{a^{\dagger2}}{2}\right]  \left \vert 0\right \rangle .
\label{11}%
\end{equation}

Correspondingly, the eigenvalue equation of $\left \vert p\right \rangle _{s,r}$
is
\begin{equation}
\left(  AP-CQ\right)  \left \vert p\right \rangle _{s,r}=p\left \vert
p\right \rangle _{s,r}, \label{12}%
\end{equation}
so $\left \vert p\right \rangle _{s,r}$ is the eigenvector of operator $AP-CQ$.
When $A=D=1,B=C=0,$the above equation evolves into $P\left \vert p\right \rangle
_{s,r}=p\left \vert p\right \rangle _{s,r},$and $\left \vert p\right \rangle
_{s,r}$ is just the momentum operator eigenstate $\left \vert p\right \rangle $.
When $A=D=0$, $-B=C=1$, the above equation evolves into $Q\left \vert
p\right \rangle _{s,r}=-p\left \vert p\right \rangle _{s,r},$and $\left \vert
p\right \rangle _{s,r}$ is just the coordinate operator eigenstate $\left \vert
x=-p\right \rangle $. So it is named the IMCS representation. $\left \vert
q\right \rangle _{s,r}$ and $\left \vert p\right \rangle _{s,r}$ will be used to
construct a new entangled state representation in the following discussion.

\section{The Intermediate coherent-entangled state (ICES)}

In this section, we propose the ICES representation and its generating protocol.

\subsection{The ICES obtained by IWOP technique}

First, let us construct the ICES by using the IWOP technique. It is well known
that the completeness relation of the coherent state $\left \vert
z\right \rangle $ is $\int \frac{d^{2}z}{\pi}\left \vert z\right \rangle
\left \langle z\right \vert =\int \frac{d^{2}z}{\pi}\colon \exp \{-(z-a)(z^{\ast
}-a^{\dagger})\} \colon=1$, which can be rewritten as%

\begin{equation}
1=\int \frac{d^{2}z}{\pi}\colon \exp \left \{  -\left(  z-\frac{a+b}{\sqrt{2}%
}\right)  \left(  z^{\ast}-\frac{a^{\dagger}+b^{\dagger}}{\sqrt{2}}\right)
\right \}  \colon, \label{13}%
\end{equation}
where $a$ and $b$ are two bosonic subtraction operators. The completeness
relation of the coordination state $\left \vert q\right \rangle $ is%

\begin{equation}
1=\int_{-\infty}^{\infty}dq\left \vert q\right \rangle \left \langle q\right \vert
=\int_{-\infty}^{\infty}\frac{dq}{\sqrt{\pi}}\colon e^{-\left(  q-Q\right)
^{2}}\colon, \label{14}%
\end{equation}
where $Q=\frac{a+a^{\dagger}}{\sqrt{2}}$is the coordinate operator. If we make
the following transform about $a^{\dagger}$ and $a$ in (\ref{14}),
\begin{equation}
a^{\dagger}\rightarrow \frac{s+r}{\sqrt{2}}\left(  b^{\dagger}-a^{\dagger
}\right)  ,a\rightarrow \frac{s^{\ast}+r^{\ast}}{\sqrt{2}}\left(  b-a\right)  ,
\label{15}%
\end{equation}
i.e.
\begin{equation}
a^{\dagger}\rightarrow \frac{\left(  D-iB\right)  }{\sqrt{2}}\left(
b^{\dagger}-a^{\dagger}\right)  ,a\rightarrow \frac{\left(  D+iB\right)
}{\sqrt{2}}\left(  b-a\right)  , \label{16}%
\end{equation}
then we have the completeness relation of the generalized coordinate state,
\begin{equation}
\frac{\pi^{-1/2}}{\sqrt{D^{2}+B^{2}}}%
%TCIMACRO{\tint _{-\infty}^{\infty}}%
%BeginExpansion
{\textstyle \int_{-\infty}^{\infty}}
%EndExpansion
dq\colon \exp \left \{  -\frac{\left[  q-\frac{\left(  D-iB\right)  \left(
b^{\dagger}-a^{\dagger}\right)  +\left(  D+iB\right)  \left(  b-a\right)  }%
{2}\right]  ^{2}}{D^{2}+B^{2}}\right \}  \colon=1. \label{17}%
\end{equation}
Then combining (\ref{13}) and (\ref{17}), we obtain the following completeness
relation%
\begin{align}
&  1=\frac{\pi^{-1/2}}{\sqrt{D^{2}+B^{2}}}\int_{-\infty}^{\infty}dq\int
\frac{d^{2}z}{\pi}\nonumber \\
&  \times \colon \exp \left \{  -\left(  z-\frac{a+b}{\sqrt{2}}\right)  \left(
z^{\ast}-\frac{a^{\dagger}+b^{\dagger}}{\sqrt{2}}\right)  \right \} \nonumber \\
&  \times \exp \left \{  -\frac{\left[  q-\frac{\left(  D-iB\right)  \left(
b^{\dagger}-a^{\dagger}\right)  +\left(  D+iB\right)  \left(  b-a\right)  }%
{2}\right]  ^{2}}{D^{2}+B^{2}}\right \}  \colon. \label{18}%
\end{align}

Moreover, by using IWOP technique and the normal product form of the two-mode
vacuum projector $\left \vert 00\right \rangle \left \langle 00\right \vert
=\colon \exp \left[  -a^{\dag}a-b^{\dag}b\right]  \colon$, the above equation
(\ref{18}) can be decomposed as%

\begin{equation}
1=\int_{-\infty}^{\infty}dq\int \frac{d^{2}z}{\pi}\left \vert z,q\right \rangle
_{s,rs,r}\left \langle z,q\right \vert , \label{19}%
\end{equation}
where $\left \vert z,q\right \rangle _{s,r}$ is called as intermediate
coherent-entangled state (ICES) and it reads as
\begin{align}
\left \vert z,q\right \rangle _{s,r}  &  =\frac{\pi^{-1/4}}{\sqrt{D+iB}}%
\exp \left \{  -\frac{\left \vert z\right \vert ^{2}}{2}-\frac{A-iC}{D+iB}%
\frac{q^{2}}{2}+\left(  \frac{z}{\sqrt{2}}+\frac{q}{D+iB}\right)  b^{\dagger
}\right \} \nonumber \\
&  \exp \left \{  \left(  \frac{z}{\sqrt{2}}-\frac{q}{D+iB}\right)  a^{\dagger
}-\frac{D-iB}{D+iB}\frac{\left(  b^{\dagger}-a^{\dagger}\right)  ^{2}}%
{4}\right \}  \left \vert 00\right \rangle \left.  \equiv \right.  \left \vert
\zeta \right \rangle . \label{20}%
\end{align}
which is the new entangled state representation. The completeness relation of
$\left \vert \zeta \right \rangle $ is just Eq.(\ref{19}).

\subsection{Generation of the ICES by using a symmetric BS and Fresnel
transform (FT)}

Next, we introduce the scheme of producing ICES through the BS and FT as
showed in Fig.1. The BS plays important role in generating some new quantum
states. For example, a variable arcsine state can be generated through a
variable BS \cite{30}, a coherent-entangled state can be generated by
asymmetric BS \cite{19}. In Fig.1, the input $a$-mode is coherent state
$\left \vert z\right \rangle _{a}$, the input $b$-mode is ICMS $\left \vert
q\right \rangle _{b,s,r}$, the two input modes are two input ports of the 50:50
BS. The initial state of the system is
\begin{equation}
\left \vert \Phi \right \rangle _{in}=\left \vert z\right \rangle _{a}%
\otimes \left \vert q\right \rangle _{b,s,r}, \label{21}%
\end{equation}

\begin{figure}[ptb]
\label{Fig 1}
\centering \includegraphics[width=0.3\columnwidth]{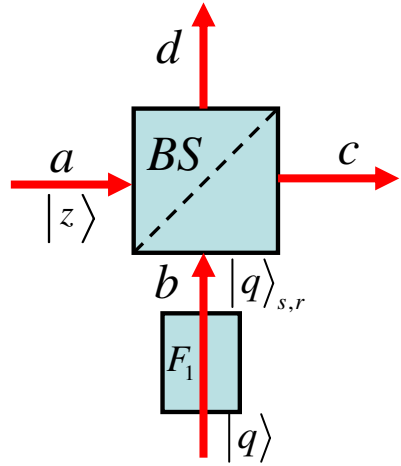}\caption{(Color
online) The project to produce the intermediate coherent-entangled state
$\left \vert \zeta \right \rangle $.}%
\end{figure}where%

\begin{equation}
\left \vert q\right \rangle _{b,s,r}=\frac{\pi^{-1/4}}{\sqrt{D+iB}}\exp \left \{
-\frac{A-iC}{D+iB}\frac{q^{2}}{2}+\frac{\sqrt{2}q}{D+iB}b^{\dagger}%
-\frac{D-iB}{D+iB}\frac{b^{\dagger2}}{2}\right \}  \left \vert 0\right \rangle
_{b}. \label{22}%
\end{equation}
According to Ref.\cite{31}, we know the beam splitter matrix transformation
between the input operators ($a^{\dagger}$, $b^{\dagger}$) and the output
operators ($c^{\dagger}$, $d^{\dagger}$),
\begin{equation}
\left(
\begin{array}
[c]{c}%
c^{\dagger}\\
d^{\dagger}%
\end{array}
\right)  =\left(
\begin{array}
[c]{cc}%
\cos \theta & \sin \theta \\
-\sin \theta & \cos \theta
\end{array}
\right)  \left(
\begin{array}
[c]{c}%
a^{\dagger}\\
b^{\dagger}%
\end{array}
\right)  =\frac{1}{\sqrt{2}}\left(
\begin{array}
[c]{cc}%
1 & 1\\
-1 & 1
\end{array}
\right)  \left(
\begin{array}
[c]{c}%
a^{\dagger}\\
b^{\dagger}%
\end{array}
\right)  , \label{23}%
\end{equation}
i.e.%

\begin{equation}
a^{\dagger}\rightarrow c^{\dagger}=\frac{1}{\sqrt{2}}\left(  a^{\dagger
}+b^{\dagger}\right)  ,b^{\dagger}\rightarrow d^{\dagger}=\frac{1}{\sqrt{2}%
}\left(  b^{\dagger}-a^{\dagger}\right)  . \label{24}%
\end{equation}

In fact, corresponding to the input and output process of the beam, the
transform process of the quantum states can be regarded as the process of an
unitary operator ($B(\theta)$) acting on a quantum state, $B(\frac{\pi}%
{4})\left \vert z\right \rangle _{a}\left \vert q\right \rangle _{b,s,r}$.
According to Ref.\cite{11a,15,33}, the BS operator $B(\theta)=e^{\theta \left(
ab^{\dagger}-a^{\dagger}b\right)  }$ satisfy the following relations
\begin{align}
B(\theta)a^{\dag}B^{^{\dagger}}  &  =a^{\dag}\cos \theta+b^{\dag}\sin
\theta,\nonumber \\
B(\theta)b^{\dag}B^{^{\dagger}}  &  =b^{\dag}\cos \theta-a^{\dag}\sin
\theta,\nonumber \\
B^{\dag}  &  =B(-\theta)=B^{-1}. \label{25}%
\end{align}
After the beam passing through the BS, the output state reads $\left \vert
\Phi \right \rangle _{out}=B(\theta)\left \vert \Phi \right \rangle _{in}$, i.e.
\begin{align}
\left \vert \Phi \right \rangle _{out}  &  =B(\frac{\pi}{4})\left \vert
z\right \rangle _{a}\otimes \left \vert q\right \rangle _{b,s,r}\nonumber \\
&  =\frac{\pi^{-1/4}}{\sqrt{D+iB}}\exp \left \{  -\frac{A-iC}{D+iB}\frac{q^{2}%
}{2}+\frac{q}{D+iB}\left(  b^{^{\dagger}}-a^{^{\dagger}}\right)  \right.
\nonumber \\
&  -\frac{D-iB}{D+iB}\frac{\left(  b^{^{\dagger}}-a^{^{\dagger}}\right)  ^{2}%
}{4}\left.  -\frac{1}{2}\left \vert z\right \vert ^{2}+z\frac{\left(
a^{^{\dagger}}+b^{^{\dagger}}\right)  }{\sqrt{2}}\right \}  \left \vert
00\right \rangle \nonumber \\
&  \equiv \left \vert z,q\right \rangle _{s,r}=\left \vert \zeta \right \rangle ,
\label{26}%
\end{align}
where we have used relations (\ref{25}). This entangled state is just the ICES
$\left \vert \zeta \right \rangle $. The whole project to produce the ICES is
showed as follow, the fist light route (route $a$) is coherent state
$\left \vert z\right \rangle _{a}$, the second light route (route $b$) is the
ICMS which can be obtained by the Fresnel transformation ($F_{1}$) from a
light in coordinate state passing through the nonlinear optical process. Then
after the two routes of light passing through the BS (linear optical
apparatus), we obtain the new entangle state ICES. So we present a new project
to produce the entangled state by mixing the BS optical apparatus and the
Fresnel transformation optical process. Thus the scheme of producing the ICES
is brought forward theoretically.

\section{The characters of the ICES}

In this section, we shall examine the characters of the ICES, including the
eigenvalue equation, orthogonal relation and Schmidt decomposition.

\subsection{ The eigenvalue equation of $\left \vert \zeta \right \rangle $}

Let annihilate operator $a$ act on the state $\left \vert \zeta \right \rangle $,
we have%
\begin{equation}
\left[  a-\frac{D-iB}{D+iB}\frac{b^{\dagger}}{2}+\frac{D-iB}{D+iB}%
\frac{a^{\dagger}}{2}\right]  \left \vert \zeta \right \rangle =\left[  \frac
{z}{\sqrt{2}}-\frac{q}{D+iB}\right]  \left \vert \zeta \right \rangle ,
\label{27}%
\end{equation}%
\begin{equation}
\left[  b+\frac{D-iB}{D+iB}\frac{b^{\dagger}}{2}-\frac{D-iB}{D+iB}%
\frac{a^{\dagger}}{2}\right]  \left \vert \zeta \right \rangle =\left[  \frac
{z}{\sqrt{2}}+\frac{q}{D+iB}\right]  \left \vert \zeta \right \rangle ,
\label{28}%
\end{equation}
Combining Eq.(\ref{27}) and (\ref{28}), we obtain the eigenvalue equations of
$\left \vert \zeta \right \rangle $, i.e.,
\begin{align}
\left(  a+b\right)  \left \vert \zeta \right \rangle  &  =\sqrt{2}z\left \vert
\zeta \right \rangle ,\nonumber \\
\left[  D\left(  Q_{b}-Q_{a}\right)  -B\left(  P_{b}-P_{a}\right)  \right]
\left \vert \zeta \right \rangle  &  =\sqrt{2}q\left \vert \zeta \right \rangle ,
\label{29}%
\end{align}
where $Q_{a}=(a+a^{\dagger})/\sqrt{2},Q_{b}=(b+b^{\dagger})/\sqrt{2}%
,P_{a}=(a-a^{\dagger})/(i\sqrt{2}),P_{b}=(b-b^{\dagger})/(i\sqrt{2}).$ It is
obvious that operators $D\left(  Q_{b}-Q_{a}\right)  -B\left(  P_{b}%
-P_{a}\right)  $ and $a+b$ are commutative, i.e., $\left[  D\left(
Q_{b}-Q_{a}\right)  -B\left(  P_{b}-P_{a}\right)  ,a+b\right]  =0$, thus the
common eigenvector of the two operators is $\left \vert \zeta \right \rangle $,
which is actually an entangled state. Especially, when $A=D=1$, $B=C=0$, the
eigenvalue equation become $\left(  Q_{b}-Q_{a}\right)  \left \vert
\zeta \right \rangle =\sqrt{2}q\left \vert \zeta \right \rangle ,\left(
a+b\right)  \left \vert \zeta \right \rangle =\sqrt{2}z\left \vert \zeta
\right \rangle .$ $\left \vert \zeta \right \rangle $ just reduces to the coherent
entangled state $\left \vert z,q\right \rangle $. When $A=D=0,-B=C=1$, the above
state $\left \vert \zeta \right \rangle $ turns to be another coherent entangled
state $\left \vert z,p\right \rangle $, whose eigenvalue equation is $\left(
P_{b}-P_{a}\right)  \left \vert \zeta \right \rangle =\sqrt{2}q\left \vert
\zeta \right \rangle ,\left(  a+b\right)  \left \vert \zeta \right \rangle
=\sqrt{2}z\left \vert \zeta \right \rangle $.

\subsection{Orthogonality of $\left \vert \zeta \right \rangle $}

Next, we shall use Eq.(\ref{26}) to obtain the orthogonality of $\left \vert
\zeta \right \rangle $. Notice that $\left \vert \zeta \right \rangle =B\left \vert
z\right \rangle _{a}\otimes \left \vert q\right \rangle _{b,s,r}$ and $\left \vert
q\right \rangle _{b,s,r}=F_{1}\left \vert q\right \rangle _{b}$, we can express
$\left \langle \zeta^{\prime}\left \vert \zeta \right \rangle \right.  $ as
\begin{align}
\left \langle \zeta^{\prime}\left \vert \zeta \right \rangle \right.   &  =\left.
_{s,r}\left \langle z^{\prime},q^{\prime}\right \vert z,q\right \rangle
_{s,r}\nonumber \\
&  =\left.  _{b}\left \langle q^{\prime}\right \vert \right.  \otimes \left.
_{a}\left \langle z^{\prime}\right \vert \right.  F_{1}^{\dag}B^{\dag}%
BF_{1}\left \vert z\right \rangle _{a}\otimes \left \vert q\right \rangle
_{b}\nonumber \\
&  =\delta \left(  q-q^{\prime}\right)  \left \langle z^{\prime}\left \vert
z\right \rangle \right. \nonumber \\
&  =\delta \left(  q-q^{\prime}\right)  \exp \left[  z^{^{\prime}\ast}z-\frac
{1}{2}\left(  \left \vert z^{\prime}\right \vert ^{2}+\left \vert z\right \vert
^{2}\right)  \right]  , \label{34}%
\end{align}
where we use the relations $B^{\dag}B=1$ and $F_{1}^{\dag}F_{1}=1$, as well as
the orthogonal relation of coordinate state $\left \langle q^{\prime}\left \vert
q\right \rangle \right.  =\delta \left(  q-q^{\prime}\right)  $. Especially,
when $z=z^{^{\prime}}$,we obtain $\left.  _{s,r}\left \langle z^{\prime
},q^{\prime}\right \vert z,q\right \rangle _{s,r}=\delta \left(  q-q^{\prime
}\right)  $, so $\left \vert z,q\right \rangle _{s,r}$ is a partially orthogonal
state, and this character is similar to the coherent state.

\subsection{The Schmidt decomposition of $\left \vert \zeta \right \rangle $}

In order to see the entanglement property of the ICES, here we consider the
Schmidt decomposing of $\left \vert \zeta \right \rangle $. Using the coordinate
representation of BS operator \cite{34},%

\begin{equation}
B(\frac{\pi}{4})=\int_{-\infty}^{\infty}dq_{1}dq_{2}\left \vert \frac{1}%
{\sqrt{2}}\left(
\begin{array}
[c]{cc}%
1 & -1\\
1 & 1
\end{array}
\right)  \left(
\begin{array}
[c]{c}%
q_{1}\\
q_{2}%
\end{array}
\right)  \right \rangle \left \langle \left(
\begin{array}
[c]{c}%
q_{1}\\
q_{2}%
\end{array}
\right)  \right \vert , \label{35}%
\end{equation}
where $\left \vert \left(
\begin{array}
[c]{c}%
q_{1}\\
q_{2}%
\end{array}
\right)  \right \rangle \equiv \left \vert q_{1}\right \rangle \otimes \left \vert
q_{2}\right \rangle $, we can obtain the Schmidt decomposing of $\left \vert
\zeta \right \rangle $,%

\begin{align}
\left \vert \zeta \right \rangle  &  =B(\frac{\pi}{4})\left \vert z\right \rangle
_{a}\left \vert q\right \rangle _{b,s,r}\nonumber \\
&  =\frac{1}{\pi^{3/4}\sqrt{2iB}}e^{-\frac{\left \vert z\right \vert ^{2}}%
{2}-\frac{z^{2}}{2}}\int_{-\infty}^{\infty}dq_{1}dq_{2}\exp \left(  \sqrt
{2}q_{1}z-\frac{1}{2}q_{1}^{2}+\frac{2q_{2}q-Dq_{2}^{2}-Aq^{2}}{2iB}\right)
\nonumber \\
&  \times \left \vert \frac{1}{\sqrt{2}}\left(  q_{1}-q_{2}\right)
\right \rangle \otimes \left \vert \frac{1}{\sqrt{2}}\left(  q_{1}+q_{2}\right)
\right \rangle , \label{36}%
\end{align}
where we used the following relations%
\begin{align}
\left.  \left \langle q_{1}\right \vert z\right \rangle _{a}  &  =\pi^{-1/4}%
\exp \left \{  -\frac{q_{1}^{2}}{2}-\frac{\left \vert z\right \vert ^{2}}{2}%
+\sqrt{2}q_{1}z-\frac{z^{2}}{2}\right \}  ,\nonumber \\
\left.  \left \langle q_{2}\right \vert q\right \rangle _{b,s,r}  &  =\frac
{\pi^{-1/2}}{\sqrt{2iB}}\exp \left[  \frac{2q_{2}q-Dq_{2}^{2}-Aq^{2}}%
{2iB}\right]  . \label{37}%
\end{align}
Let$\frac{q_{1}-q_{2}}{\sqrt{2}}=q^{\prime},$then $q_{1}=\sqrt{2}q^{\prime
}+q_{2}$. Substituting the expression of $q_{1}$ into (\ref{36}), we obtain%
\begin{align}
\left \vert \zeta \right \rangle  &  =\frac{1}{\pi^{3/4}\sqrt{iB}}\int_{-\infty
}^{\infty}dq^{\prime}dq_{2}\left \vert q^{\prime}\right \rangle _{a}%
\otimes \left \vert \left(  q^{\prime}+\sqrt{2}q_{2}\right)  \right \rangle
_{b}\exp \left(  \frac{2q_{2}q-Dq_{2}^{2}-Aq^{2}}{2iB}\right) \nonumber \\
&  \times \exp \left[  -\frac{\left \vert z\right \vert ^{2}}{2}+\left(
2q^{\prime}+\sqrt{2}q_{2}\right)  z-\frac{z^{2}}{2}-\frac{1}{2}\left(
\sqrt{2}q^{\prime}+q_{2}\right)  ^{2}\right]  , \label{38}%
\end{align}
which is the Schmidt decomposing of $\left \vert \zeta \right \rangle $ in
coordinate representation. Thus $\left \vert \zeta \right \rangle $ is just an
entangled state.

\subsection{Conjugate state of $\left \vert \zeta \right \rangle $}

We can also obtain the conjugate state $\left \vert \kappa \right \rangle $ of
$\left \vert \zeta \right \rangle $ by\ considering the input mode $a$ as
coherent state $\left \vert z\right \rangle $, the input mode $b$ as the IMCS
$\left \vert p\right \rangle _{s,r}$. Considering that the input $a$-mode is
coherent state $\left \vert z\right \rangle $, the input $b$-mode is the IMCS
$\left \vert p\right \rangle _{s,r}$, the two input modes are two input ports of
the Beam splitter (BS), the ratio between the reflection coefficient and the
transmission coefficient is $50\colon50$. The initial state of the system is
$\left \vert \Phi \right \rangle _{in}=\left \vert z\right \rangle _{a}%
\otimes \left \vert p\right \rangle _{b,s,r},$ where
\begin{equation}
\left \vert p\right \rangle _{b,s,r}=\frac{\pi^{-1/4}}{\sqrt{A-iC}}\exp \left[
-\frac{D+iB}{2\left(  A-iC\right)  }p^{2}+\frac{\sqrt{2}ip}{A-iC}b^{\dagger
}+\frac{A+iC}{2\left(  A-iC\right)  }b^{\dagger2}\right]  \left \vert
0\right \rangle . \label{39}%
\end{equation}
After the beam passing through the BS, in a similar way we have
\begin{align}
\left \vert \kappa \right \rangle  &  =B(\frac{\pi}{4})\left \vert z\right \rangle
_{a}\otimes \left \vert p\right \rangle _{b,s,r}\nonumber \\
&  =\frac{\pi^{-1/4}}{\sqrt{A-iC}}\exp \left \{  -\frac{\left \vert z\right \vert
^{2}}{2}+z\frac{\left(  a^{\dagger}+b^{\dagger}\right)  }{\sqrt{2}}\right \}
\nonumber \\
&  \times \exp \left \{  -\frac{D+iB}{A-iC}\frac{p^{2}}{2}+\frac{ip}{A-iC}\left(
b^{\dagger}-a^{\dagger}\right)  +\frac{A+iC}{\left(  A-iC\right)  }%
\frac{\left(  b^{\dagger}-a^{\dagger}\right)  ^{2}}{4}\right \}  \left \vert
0\right \rangle _{a}\left \vert 0\right \rangle _{b}. \label{40}%
\end{align}
This new representation is another kind of the intermediate coherent-entangled
state $\left \vert \kappa \right \rangle $. The eigenvalue equation of the state
$\left \vert \kappa \right \rangle $ are%
\begin{align}
\left(  b+a\right)  \left \vert \kappa \right \rangle  &  =\sqrt{2}z\left \vert
\kappa \right \rangle ,\nonumber \\
\left[  A\left(  P_{b}-P_{a}\right)  -C\left(  Q_{b}-Q_{a}\right)  \right]
\left \vert \kappa \right \rangle  &  =\sqrt{2}p\left \vert \kappa \right \rangle .
\label{41}%
\end{align}
Thus $\left \vert \kappa \right \rangle $ is the common eigenvector of $a+b$ and
$A\left(  P_{b}-P_{a}\right)  -C\left(  Q_{b}-Q_{a}\right)  $. We can easily
obtain $[A\left(  P_{b}-P_{a}\right)  -C\left(  Q_{b}-Q_{a}\right)  ,D\left(
Q_{b}-Q_{a}\right)  -B\left(  P_{b}-P_{a}\right)  ]$=$-2i\hbar \left(
AD-BC\right)  =-2i\hbar,$so $\left \vert \kappa \right \rangle $ is the conjugate
state of $\left \vert \zeta \right \rangle $.

\begin{figure}[ptb]
\label{Fig 2}
\centering \includegraphics[width=0.3\columnwidth]{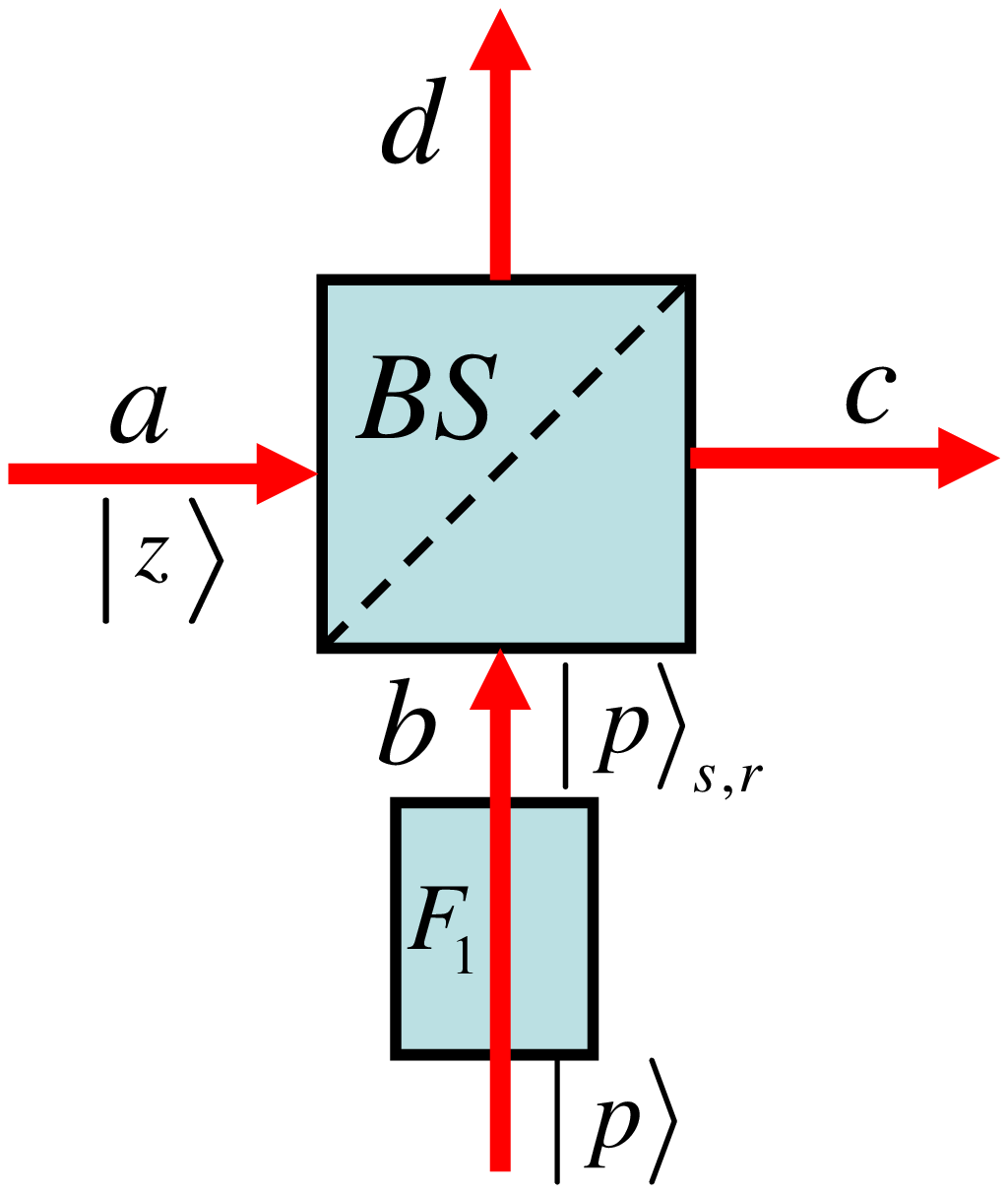}\caption{(Color
online) The project to produce the intermediate coherent-entangled state
$\left \vert \kappa \right \rangle $.}%
\end{figure}

\section{Applications of the ICES}

\subsection{To obtain the squeezing operator and its exponential form}

As an important application of $\left \vert z,q\right \rangle _{s,r}$, we derive
a new squeezing operator. By constructing the following ket-bra integration,
\begin{equation}
U=\frac{1}{\sqrt{\mu}}\int_{-\infty}^{\infty}dq\int \frac{d^{2}z}{\pi
}\left \vert z,\frac{q}{\mu}\right \rangle _{r,sr,s}\left \langle z,q\right \vert
, \label{43}%
\end{equation}
where $\mu=e^{\lambda}$ , $\left \vert z,\frac{q}{\mu}\right \rangle
_{s,r}=B\left \vert z\right \rangle _{a}\otimes \left \vert \frac{q}{\mu
}\right \rangle _{b,s,r}$ , $\left \vert \frac{q}{\mu}\right \rangle
_{b,s,r}=F_{1}\left \vert \frac{q}{\mu}\right \rangle _{b}$, and $\left \vert
\frac{q}{\mu}\right \rangle _{b}=\sqrt{\mu}S_{1}\left \vert q\right \rangle _{b}%
$, in which $S_{1}\left(  \lambda \right)  =e^{\frac{\lambda}{2}\left(
b^{2}-b^{\dagger2}\right)  }$ is the single mode squeezing operator, $F_{1}$
is the Fresnel operator which acts on mode $b$ only, $B$($\frac{\pi}{4}$) is
the BS operator which acts on mode $a$ and mode $b$ simultaneously. Thus we
have
\begin{align}
\left \vert z,\frac{q}{\mu}\right \rangle _{s,r}  &  =BF_{1}\left \vert
z\right \rangle _{a}\otimes \left \vert \frac{q}{\mu}\right \rangle _{b}=\sqrt
{\mu}BF_{1}S_{1}\left \vert z\right \rangle _{a}\left \vert q\right \rangle
_{b},\nonumber \\
\left \langle z,q\right \vert _{r,s}  &  =\left.  _{a}\left \langle z\right \vert
_{b}\left \langle q\right \vert \right.  F_{1}^{\dag}B^{\dag}, \label{44}%
\end{align}
Using the following relations
\begin{align}
F_{1}(s,r)bF_{1}^{\dagger}(s,r)  &  =s^{\ast}b+rb^{\dagger},F_{1}^{\dagger
}(s,r)bF_{1}(s,r)=sb-rb^{\dagger},\nonumber \\
F_{1}(s,r)b^{\dagger}F_{1}^{\dagger}(s,r)  &  =sb^{\dagger}+r^{\ast}%
b,F_{1}^{\dagger}(s,r)b^{\dagger}F_{1}(s,r)=s^{\ast}b^{\dagger}-r^{\ast
}b,\nonumber \\
F_{1}^{\dag}  &  =F_{1}\left(  r\rightarrow-r,s\rightarrow s^{\ast}\right)
,\nonumber \\
S_{1}bS_{1}^{\dag}  &  =b\cosh \lambda+b^{\dag}\sinh \lambda, \label{45}%
\end{align}
and the completeness relations ($\int_{-\infty}^{\infty}dq\int \frac{d^{2}%
z}{\pi}\left \vert z,q\right \rangle \left \langle z,q\right \vert =1$), we
obtain
\begin{align}
U  &  =\frac{1}{\sqrt{\mu}}\int_{-\infty}^{\infty}dq\int \frac{d^{2}z}{\pi
}\left \vert z,\frac{q}{\mu}\right \rangle _{s,rs,r}\left \langle z,q\right \vert
\nonumber \\
&  =BF_{1}S_{1}\int_{-\infty}^{\infty}dq\int \frac{d^{2}z}{\pi}\left \vert
z,q\right \rangle \left \langle z,q\right \vert F_{1}^{\dag}B^{\dag}\nonumber \\
&  =BF_{1}S_{1}F_{1}^{\dagger}B^{\dagger}\nonumber \\
&  =\exp \left \{  \frac{\lambda}{2}\left[  \frac{1}{2}\left(  s^{\ast2}%
-r^{\ast2}\right)  \left(  a-b\right)  ^{2}\right]  \right. \nonumber \\
&  \left.  +\frac{\lambda}{2}\left[  s^{\ast}r\left(  1+b^{\dagger
}b+a^{\dagger}a-a^{\dagger}b-ab^{\dagger}\right)  \right]  -c.c\right \}  ,
\label{46}%
\end{align}
where $c.c$ represents Hermitian conjugate. It is obvious that $U^{\dag
}=U\left(  -\lambda \right)  =U^{-1}$ is an unitary\ operator. The last
equation of Eq.(\ref{46}) is the exponential form of unitary $U$. Using the
Baker-Hausdroff formula $e^{\xi A}Be^{-\xi A}=B+\xi \left[  A,B\right]
+\frac{\xi^{2}}{2!}\left[  A,\left[  A,B\right]  \right]  +...$, we can get%
\begin{align}
&
\begin{array}
[c]{c}%
UaU^{\dagger}=\frac{1}{2}[\left(  b-a\right)  \left[  \left(  rs^{\ast
}-sr^{\ast}\right)  \sinh \lambda-\cosh \lambda \right] \\
+\left(  b^{\dagger}-a^{\dagger}\right)  \left(  r^{2}-s^{2}\right)
\sinh \lambda+\left(  a+b\right)  ],
\end{array}
\nonumber \\
&
\begin{array}
[c]{c}%
UbU^{\dagger}=\frac{1}{2}[\left(  b-a\right)  \left[  \cosh \lambda-\left(
rs^{\ast}-sr^{\ast}\right)  \sinh \lambda \right] \\
+\left(  b^{\dagger}-a^{\dagger}\right)  \left(  s^{2}-r^{2}\right)
\sinh \lambda+\left(  a+b\right)  ],
\end{array}
\label{47}%
\end{align}
where we have used Eq. (\ref{25}) and the relation $ss^{\ast}-rr^{\ast}=1$. It
then follows that%

\begin{align}
&
\begin{array}
[c]{c}%
UQ_{a}U^{\dagger}=\frac{1}{2}\{ \left(  Q_{a}+Q_{b}\right)  +i\left(
P_{a}-P_{b}\right)  \frac{\left(  r-s^{\ast}\right)  ^{2}-\left(  r^{\ast
}-s\right)  ^{2}}{2}\sinh \lambda \\
+\left(  Q_{a}-Q_{b}\right)  [\cosh \lambda-\frac{r^{2}-s^{2}+r^{\ast2}%
-s^{\ast2}}{2}\sinh \lambda]\},
\end{array}
\nonumber \\
&
\begin{array}
[c]{c}%
UQ_{b}U^{\dagger}=\frac{1}{2}\{ \left(  Q_{a}+Q_{b}\right)  -i\left(
P_{a}-P_{b}\right)  \frac{\left(  r-s^{\ast}\right)  ^{2}-\left(  r^{\ast
}-s\right)  ^{2}}{2}\sinh \lambda \\
+\left(  Q_{b}-Q_{a}\right)  [\cosh \lambda-\frac{r^{2}-s^{2}+r^{\ast2}%
-s^{\ast2}}{2}\sinh \lambda]\},
\end{array}
\label{48}%
\end{align}
and
\begin{align}
&
\begin{array}
[c]{c}%
UP_{a}U^{\dagger}=\frac{1}{2}\{ \left(  P_{a}+P_{b}\right)  +\left(
Q_{b}-Q_{a}\right)  \frac{\left(  r+s^{\ast}\right)  ^{2}-\left(  r^{\ast
}+s\right)  ^{2}}{2i}\sinh \lambda \\
+\left(  P_{b}-P_{a}\right)  [\frac{s^{\ast2}-r^{\ast2}-r^{2}+s^{2}}{2}%
\sinh \lambda-\cosh \lambda]\},
\end{array}
\nonumber \\
&
\begin{array}
[c]{c}%
UP_{b}U^{\dagger}=\frac{1}{2}\{ \left(  P_{a}+P_{b}\right)  +\left(
Q_{b}-Q_{a}\right)  \frac{\left(  r^{\ast}+s\right)  ^{2}-\left(  r+s^{\ast
}\right)  ^{2}}{2i}\sinh \lambda \\
+\left(  P_{b}-P_{a}\right)  [\frac{-s^{\ast2}+r^{\ast2}-s^{2}+r^{2}}{2}%
\sinh \lambda+\cosh \lambda]\}.
\end{array}
\label{49}%
\end{align}
Furthermore, combining Eq. (\ref{48}) and (\ref{49}), we obtain%
\begin{align}
U\left(  Q_{a}+Q_{b}\right)  U^{\dagger}  &  =Q_{a}+Q_{b},\nonumber \\
U\left(  Q_{a}-Q_{b}\right)  U^{\dagger}  &  =\{i\left(  P_{a}-P_{b}\right)
\frac{\left(  r-s^{\ast}\right)  ^{2}-\left(  r^{\ast}-s\right)  ^{2}}{2}%
\sinh \lambda \nonumber \\
&  +\left(  Q_{a}-Q_{b}\right)  [\cosh \lambda-\frac{\left(  r^{2}%
-s^{2}+r^{\ast2}-s^{\ast2}\right)  }{2}\sinh \lambda]\},\nonumber \\
U\left(  P_{a}+P_{b}\right)  U^{\dagger}  &  =P_{a}+P_{b},\nonumber \\
U\left(  P_{a}-P_{b}\right)  U^{\dagger}  &  =\{-i\left(  Q_{b}-Q_{a}\right)
\frac{\left(  r+s^{\ast}\right)  ^{2}-\left(  r^{\ast}+s\right)  ^{2}}{2}%
\sinh \lambda \nonumber \\
&  +\left(  P_{b}-P_{a}\right)  [\frac{s^{\ast2}-r^{\ast2}-r^{2}+s^{2}}%
{2}\sinh \lambda-\cosh \lambda]\}. \label{50}%
\end{align}
$\  \  \  \  \  \  \  \  \  \  \  \  \  \  \  \  \  \  \  \  \  \  \  \ $

When $s=1,r=0$, the above results just degenerate to the case of two-mode
squeezing operator $U=\frac{1}{\sqrt{\mu}}\int_{-\infty}^{\infty}dq\int
\frac{d^{2}z}{\pi}\left \vert z,\frac{q}{\mu}\right \rangle \left \langle
z,q\right \vert $ (see reference \cite{18}). If we define $a^{\prime \dagger
}=Ua^{\dagger}U^{\dagger}$, we can verify $\left[  a^{\prime},a^{\prime
\dagger}\right]  =1$.

\subsection{To derive some operator identities}

In quantum mechanics, quantum operators are usually non-commuting, as many as
possible operator identities are searched to apply in all kinds of physical
problems. By using the intermediate coherent-entangled state $\left \vert
\zeta \right \rangle $ and the IWOP technique, we can construct some operator
identities, such as $e^{D\left(  Q_{b}-Q_{a}\right)  -B\left(  P_{b}%
-P_{a}\right)  }e^{a+b}=\colon?\colon$, and $\exp \left \{  -y\left[  D\left(
Q_{b}-Q_{a}\right)  -B\left(  P_{b}-P_{a}\right)  \right]  ^{2}\right \}
=\colon?\colon$.

In order to obtain the normally ordered expansion of $e^{D\left(  Q_{b}%
-Q_{a}\right)  -B\left(  P_{b}-P_{a}\right)  }e^{a+b}$, by noticing the
eigenvalue equation of state $\left \vert \zeta \right \rangle $ (Eq.(\ref{29}))
and the completeness relation of the state $\left \vert \zeta \right \rangle $
(Eq.(\ref{19})) as well as the expression of state $\left \vert \zeta
\right \rangle $ (Eq.(\ref{20})), we have the operator identity
\begin{align}
&  e^{D\left(  Q_{b}-Q_{a}\right)  -B\left(  P_{b}-P_{a}\right)  }%
e^{a+b}\nonumber \\
&  =\int_{-\infty}^{\infty}dq\int \frac{d^{2}z}{\pi}e^{\sqrt{2}q}e^{\sqrt{2}%
z}\left \vert z,q\right \rangle _{s,rs,r}\left \langle z,q\right \vert \nonumber \\
&  =\colon \exp \left \{  \left[  D\left(  Q_{b}-Q_{a}\right)  -B\left(
P_{b}-P_{a}\right)  \right]  +\frac{D^{2}+B^{2}}{2}+a+b\right \}  \colon,
\label{53}%
\end{align}
where we have used the IWOP technique and the integration formulas
\begin{equation}
\int \frac{d^{2}z}{\pi}e^{\xi \left \vert z\right \vert ^{2}+\eta z+\gamma
z^{\ast}}=-\frac{1}{\xi}\exp \left[  -\frac{\eta \gamma}{\xi}\right]  ,
\label{54}%
\end{equation}
and
\begin{equation}
\int_{-\infty}^{\infty}\exp \left(  -\alpha x^{2}+\beta x\right)
dx=\sqrt{\frac{\pi}{\alpha}}\exp \left(  \frac{\beta^{2}}{4\alpha}\right)  .
\label{55}%
\end{equation}

Similarly, by using Eqs.(\ref{40},\ref{41}), we also have%
\begin{align}
&  e^{A\left(  P_{b}-P_{a}\right)  -C\left(  Q_{b}-Q_{a}\right)  }%
e^{a+b}\nonumber \\
&  =\int_{-\infty}^{\infty}dp\int \frac{d^{2}z}{\pi}e^{A\left(  P_{b}%
-P_{a}\right)  -C\left(  Q_{b}-Q_{a}\right)  }e^{a+b}\left \vert \kappa
_{2}\right \rangle \left \langle \kappa_{2}\right \vert \nonumber \\
&  =\int_{-\infty}^{\infty}dp\int \frac{d^{2}z}{\pi}e^{\sqrt{2}p}e^{\sqrt{2}%
z}\left \vert \kappa_{2}\right \rangle \left \langle \kappa_{2}\right \vert
\nonumber \\
&  =\colon \exp \left[  a+b+\frac{\left(  A^{2}+C^{2}\right)  }{2}+A\left(
P_{b}-P_{a}\right)  -C\left(  Q_{b}-Q_{a}\right)  \right]  \colon, \label{56}%
\end{align}
and
\begin{align}
&  \exp \left \{  -y\left[  D\left(  Q_{b}-Q_{a}\right)  -B\left(  P_{b}%
-P_{a}\right)  \right]  ^{2}\right \} \nonumber \\
&  =\int_{-\infty}^{\infty}dq\int \frac{d^{2}z}{\pi}e^{-y\left(  \sqrt
{2}q\right)  ^{2}}\left \vert z,q\right \rangle _{s,rs,r}\left \langle
z,q\right \vert \nonumber \\
&  =\colon \sqrt{\frac{1}{1+2y\left(  D^{2}+B^{2}\right)  }}\exp \left \{
\frac{-y\left[  D\left(  Q_{b}-Q_{a}\right)  -B\left(  P_{b}-P_{a}\right)
\right]  ^{2}}{1+2y\left(  D^{2}+B^{2}\right)  }\right \}  \colon, \label{57}%
\end{align}
as well as
\begin{align}
&  \left[  D\left(  Q_{b}-Q_{a}\right)  -B\left(  P_{b}-P_{a}\right)  \right]
^{n}\nonumber \\
&  =\int_{-\infty}^{\infty}dq\int \frac{d^{2}z}{\pi}\left[  D\left(
Q_{b}-Q_{a}\right)  -B\left(  P_{b}-P_{a}\right)  \right]  ^{n}\left \vert
z,q\right \rangle _{s,r,s,r}\left \langle z,q\right \vert \nonumber \\
&  =\int_{-\infty}^{\infty}dq\int \frac{d^{2}z}{\pi}\left[  \sqrt{2}q\right]
^{n}\left \vert z,q\right \rangle _{s,r,s,r}\left \langle z,q\right \vert
\nonumber \\
&  =\colon \left(  i\sqrt{\frac{D^{2}+B^{2}}{2}}\right)  ^{n}H_{n}%
(\frac{D\left(  Q_{b}-Q_{a}\right)  -B\left(  P_{b}-P_{a}\right)  }{i\sqrt
{2}\sqrt{D^{2}+B^{2}}})\colon, \label{58}%
\end{align}
where we have used the generating function of Hermite polynomial
\begin{equation}
H_{n}(x)=\left.  \frac{\partial^{n}}{\partial t^{n}}\exp \left(  2xt-t^{2}%
\right)  \right \vert _{t=0}. \label{59}%
\end{equation}
Thus we can get some operator identities shown in Eqs.(\ref{53})-(\ref{59}).
These operator identities can be used in the quantum mechanics problems.

\section{Conclusion}

In summary, we have proposed a new entangled state representation (the ICES
representation $\left \vert z,q\right \rangle _{s,r}$) naturally from the
completeness relations of the coherent state and the generalized coordinate
state. We also have shown that mixing a symmetric beam splitter and the
Fresnel transformation optical process can generate this new bipartite
entangled state $\left \vert z,q\right \rangle _{s,r}$, which is a new project.
We also discuss the new state's properties, i.e. the partly orthogonality and
the eigenvalue equation. $\left \vert z,q\right \rangle _{s,r}$ , also provides
a way to predict a new squeezing operator and some new operator identities.
These discussions not only exhibit the usefulness of the ICES representation,
but also provides considerable insight into the character of them.

\begin{acknowledgments}
This project was supported by the National Natural Science Foundation of China
(Nos.11264018, 11364022, 11464018), the Natural Science Foundation of Jiangxi
Province (Nos. 20142BAB202004 and 20151BAB212006), and the Research Foundation
of the Education Department of Jiangxi Province of China (No.GJJ12171,
GJJ14274), as well as Degree and postgraduate education teaching reform
project of jiangxi province (No. JXYJG-2013-027).
\end{acknowledgments}

\end{document}